\documentclass[letterpaper,english,reprint, aps]{revtex4-1}
\usepackage[T1]{fontenc}
\usepackage[latin9]{inputenc}
\usepackage{array}
\usepackage{multirow}
\usepackage{amsmath}
\usepackage{amssymb}
\usepackage{graphicx}

\makeatletter

\pdfpageheight\paperheight
\pdfpagewidth\paperwidth

\providecommand{\tabularnewline}{\\}

\makeatother

\usepackage{babel}
\begin{document}
\preprint{APS/123-QED}
\title{Performance improvement of a fractional quantum Stirling heat engine}
\author{Shihao Xia}
\affiliation{Department of Physics, Xiamen University, Xiamen 361005, People's
Republic of China}
\author{Youlin Wang}
\affiliation{Department of Physics, Xiamen University, Xiamen 361005, People's
Republic of China}
\author{Minglong Lv}
\affiliation{Department of Physics, Xiamen University, Xiamen 361005, People's
Republic of China}
\author{Jincan Chen}
\affiliation{Department of Physics, Xiamen University, Xiamen 361005, People's
Republic of China}
\author{Shanhe Su}
\email{sushanhe@xmu.edu.cn}

\affiliation{Department of Physics, Xiamen University, Xiamen 361005, People's
Republic of China}
\date{\today}
\begin{abstract}
To investigate the impact of fractional parameter on the thermodynamic
behaviors of quantum systems, we incorporate fractional quantum mechanics
into the cycle of a quantum Stirling heat engine and examine the influence
of fractional parameter on the regeneration and efficiency. We propose
a novel approach to control the thermodynamic cycle that leverages
the fractional parameter structure and evaluates its effectiveness.
Our findings reveal that by tuning the fractional parameter, the region
of the cycle with the perfect regeneration and the Carnot efficiency
can be expanded.
\end{abstract}
\maketitle

\section{\label{sec:leveli}Introduction}

The study of fractional calculus \citep{herrmann2011fractional,kilbas1993fractional,butzer2000introduction}
has received growing attention in recent years due to its unique mathematical
structure and close association with the renormalization and the inverse
power law. It provides a powerful mathematical tool to solve problems
related to complex systems \citep{west2014colloquium,guo2021renormalization}.
In addition, Lévy flight, a natural generalization of Brownian motion,
has become a research hotspot in the field of anomalous diffusion
with practical implications for the advancements of physics, life
science, information science, and other disciplines \citep{khinchine1936lois,mandelbrot1982fractal,de2011levy,zaburdaev2015levy,barthelemy2008levy,margolin2005nonergodicity,liu2016dynamical}.
Lévy flight arises from the strong interaction between particles and
their environment, and it is a Markov stochastic process characterized
by long-range jumps. Although the Lévy process is mainly utilized
for numerical simulations, the experimental work \citep{sagi2012observation}
has shown that it is feasible to adjust the system parameters with
precision, enabling direct experimental studies of Lévy flight. Hence,
discussions on the diffusion behavior and the dynamics of atomic groups
with damping and even more complex transport environments are underway.

Applications of fractional quantum mechanics have been developed by
defining the fractional path integral over Lévy paths and using the
Riesz fractional derivative, extending the concept of fractality in
quantum physics \citep{laskin2000,laskin2000fractals,laskin2000fractional,laskin2002fractional,book93354028}.
This area has witnessed significant advances in recent years \citep{naber2004time,wang2007generalized,dong2008space,felmer2012positive,secchi2013ground,fall2015ground,wei2015some,longhi2015fractional,laskin2017time},
and has been demonstrated experimentally \citep{liu2023experimental}.
Moreover, fractional calculus is increasingly being employed to describe
thermodynamic phenomena \citep{mainardi1997fractional,tudor2007statistical,wang2014fractional,meilanov2014thermodynamics,bagci2016third,lopes2020review,tarasov2006fractional,sisman2021fractional,khadem2022stochastic,korichi2022thermal}.
Attempts have also been made to combine fractional quantum mechanics
with thermodynamics, such as Black hole thermodynamics  \citep{jalalzadeh2021prospecting},
thermal properties of fractional quantum Dirac oscillators \citep{korichi2022thermal},
and etc.

Quantum heat engine \citep{scovil1959three,geusic1967quantum,quan2005quantum,kieu2006quantum,quan2007quantum,wu1998performance,huang2014quantum}
is an excellent platform for studying the thermodynamic properties
of quantum systems. In this context, we investigate the effect of
the fractional parameter on the performance of a quantum Stirling
engine (QSE). We propose a new thermodynamic process based on the
fractional parameter and analyze the behavior of the thermodynamic
cycle that incorporate this process. It will demonstrate the potential
applications of fractional quantum mechanics in thermodynamics. 

This paper is organized as follows: In Section \ref{sec:levelii},
we provide a brief overview of fractional quantum mechanics and show
the solution in the infinite potential well (IPW). Several fundamental
concepts of quantum thermodynamics are introduced as well. In Section
\ref{sec:QUANTUM-STIRLING-ENGINE}, we introduce the structure of
the QSE and propose a new way to regulate the thermodynamic cycle
based on fractional parameters. Expressions of thermodynamic quantities
in the cycle are provided. In Section \ref{sec:result-and-discussion},
the effects of the fractional parameter on the performance of the
QSE are discussed. Conclusions are given in Section \ref{sec:conclusion}. 

\section{\label{sec:levelii}FRACTIONAL QUANTUM MECHANICS AND KEY quantities
IN QUANTUM THERMODYNAMIC PROCESSES}

\subsection{FRACTIONAL QUANTUM MECHANICS}

In fractional quantum mechanics, the fractional Hamiltonian operator
is defined as $H=\ensuremath{D_{\alpha}|p|^{\alpha}+V(x)}\text{,}$
where $p$ is the momentum, the fractional parameter $1<\alpha\leq2$,
$V(x)$ is the potential energy as a functional of a particle path
$x$ , and $D_{\alpha}$ is the scale coefficient \citep{laskin2000,korichi2022thermal}.
If the system at an initial time $t_{a}$ starts from the point $x_{a}$
and goes to the final point $x_{b}$ at time $t_{b}$, one could define
the quantum-mechanical amplitude, often called a kernel, $K\left(x_{b}t_{b}\mid x_{a}t_{a}\right)$.
The kernel function is the sum of the contributions of all trajectories
through the first and last points \citep{laskin2000,laskin2000fractals,laskin2000fractional,laskin2002fractional,book93354028}.
The kernel based on the Lévy path in phase space is defined as

\begin{equation}
\begin{aligned} & K\left(x_{b}t_{b}\mid x_{a}t_{a}\right)=\lim_{N\rightarrow\infty}\int_{-\infty}^{\infty}dx_{1}\ldots dx_{N-1}\frac{1}{(2\pi\hbar)^{N}}\\
 & \times\int_{-\infty}^{\infty}dp_{1}\ldots dp_{N}\exp\left\{ \frac{i}{\hbar}\sum_{j=1}^{N}p_{j}\left(x_{j}-x_{j-1}\right)\right\} \\
 & \times\exp\left\{ -\frac{i}{\hbar}D_{\alpha}\varepsilon\sum_{j=1}^{N}\left|p_{j}\right|^{\alpha}-\frac{i}{\hbar}\varepsilon\sum_{j=1}^{N}V\left(x_{j}\right)\right\} ,
\end{aligned}
\label{eq:kernel}
\end{equation}
where $\hbar$ is Planck\textquoteright s constant, $\varepsilon=\left(t_{b}-t_{a}\right)/N$,
$x_{j}=x\left(t_{a}+j\varepsilon\right)$, $p_{j}=p\left(t_{a}+j\varepsilon\right)$,
$x\left(t_{a}+j\varepsilon\right)_{j=0}=x_{a}$, and $x\left(t_{a}+j\varepsilon\right)_{j=N}=x_{b}$.

The kernel describes the evolution of a system, leading to the fractional
wave function at time $t_{b}$ 

\begin{equation}
\psi\left(x_{b},t_{b}\right)=\int_{-\infty}^{\infty}dx_{a}K\left(x_{b}t_{b}\mid x_{a}t_{a}\right)\psi\left(x_{a},t_{a}\right),\label{eq:integralK}
\end{equation}
with $\psi\left(x_{a},t_{a}\right)$ being the fractional wave function
of the initial state. The fractional wave function $\psi\left(x,t\right)$
satisfies the fractional Schrödinger equation (Appendix A)

\begin{equation}
i\hbar\frac{\partial\psi(x,t)}{\partial t}=-D_{\alpha}(\hbar\nabla)^{\alpha}\psi(x,t)+V(x)\psi(x,t),\label{eq:FSeq}
\end{equation}
where the quantum Riesz fractional derivative $(\hbar\nabla)^{\alpha}$
is defined as

\begin{equation}
(\hbar\nabla)^{\alpha}\psi(x,t)=-\frac{1}{2\pi\hbar}\int_{-\infty}^{\infty}dp\:\text{exp}\left(i\frac{px}{\hbar}\right)|p|^{\alpha}\varphi(p,t)\label{eq:RieszD}
\end{equation}
with $\varphi(p,t)=\int_{-\infty}^{\infty}dp\:\text{exp}\left(-i\frac{px}{\hbar}\right)\psi(x,t)$
being the Fourier transform of $\psi(x,t)$. 

In the following discussion, the scale coefficient $D_{\alpha}$ is
set to be equal to $\left(1/2m\right)^{\frac{\alpha}{2}}$ with $m$
being the mass of the quantum mechanical particle \citep{korichi2022thermal}.
For $\alpha=2$, it becomes the standard quantum mechanics that we
know. Meanwhile, we consider a particle in a one-dimensional IPW,
where the potential field

\begin{equation}
\ensuremath{V(x)=\begin{cases}
0 & -L/2\leqslant x\leqslant L/2,\\
\infty & \mathrm{otherwise.}
\end{cases}}\label{eq:potential}
\end{equation}

The solution of Eq. (\ref{eq:FSeq}) is related to the time independent
wave function $\phi(x)$ by

\begin{equation}
\psi(x,t)=\exp\left\{ -i\frac{Et}{\hbar}\right\} \phi(x),\label{eq:finx}
\end{equation}
where $E$ represents the energy of the particle. Putting Eq. (\ref{eq:finx})
into Eq. (\ref{eq:FSeq}) leads to the following time-independent
fractional Schrödinger equation 

\begin{equation}
-D_{\alpha}(\hbar\nabla)^{\alpha}\phi(x)+V(x)\phi(x)=E\phi(x).\label{eq:TIFSeq}
\end{equation}

By using Eqs. (\ref{eq:potential}) and (\ref{eq:TIFSeq}) and considering
the boundary conditions, the eigenvalue $E_{n}\left(L,\alpha\right)$
of the fractional Hamiltonian operator $H$ and the corresponding
wave function $\phi(x)$ read \citep{laskin2000fractals}

\begin{equation}
\begin{aligned}E_{n} & \left(L,\alpha\right)=D_{\alpha}\left(\frac{2\pi\hbar}{L}\right)^{\alpha}n^{\alpha}\\
 & =\left(\frac{1}{2m}\right)^{\frac{\alpha}{2}}\left(\frac{2\pi\hbar}{L}\right)^{\alpha}n^{\alpha},
\end{aligned}
\label{eq:IPWenergy}
\end{equation}

\begin{equation}
\phi(x)=\begin{cases}
\sqrt{\frac{2}{L}}\cos\left[\left(n-\frac{1}{2}\right)\frac{2\pi x}{L}\right] & \mathrm{for\:\mathit{n\:}even\text{,}}\\
\sqrt{\frac{2}{L}}\sin\frac{2n\pi x}{L} & \mathrm{for\:}n\:\mathrm{odd},
\end{cases}
\end{equation}
where $L$ represents the width of the potential well, and $n$ is
a positive integer $(n=1,2,3,4,...)$.

\subsection{KEY QUANTITIES IN QUANTUM THERMODYNAMIC PROCESSES}

The internal energy $U$ of the particle is expressed as the ensemble
average of the fractional Hamiltonian operator, i.e.,

\begin{equation}
U=\langle H\rangle=\sum_{n}P_{n}E_{n},\label{eq:U}
\end{equation}
where $P_{n}$ denotes the occupation probability of the $n$th eigenstate
with energy $E_{n}$. During an infinitesimal process, the time differential
of the internal energy

\begin{equation}
dU=\sum_{n}\left(E_{n}dP_{n}+P_{n}dE_{n}\right).
\end{equation}

According to the first law of thermodynamics, $dU$ is associated
with the heat $\mkern3mu\mathchar'26\mkern-12mu dQ$ absorbed from
the environment and the work $\mkern3mu\mathchar'26\mkern-12mu dW$
performed by the external agent, i.e.,

\begin{equation}
dU=\mkern3mu\mathchar'26\mkern-12mu dQ+\mkern3mu\mathchar'26\mkern-12mu dW.
\end{equation}

For the isothermal and isochoric processes, the heat exchange and
the work done during an infinitesimal thermodynamic process are, respectively,
identified as \citep{quan2005quantum,kieu2006quantum,quan2007quantum,su2018heat} 

\begin{equation}
\mkern3mu\mathchar'26\mkern-12mu dQ=\sum_{n}E_{n}dP_{n},\label{eq:dQ}
\end{equation}
and

\begin{equation}
\mkern3mu\mathchar'26\mkern-12mu dW=\sum_{n}P_{n}dE_{n}.\label{eq:dW}
\end{equation}

As the isothermal process with the temperature $T$ of the particle
being a constant is reversible, Eq. (\ref{eq:dQ}) is equivalent to
\begin{equation}
\mkern3mu \mkern3mu\mathchar'26\mkern-12mu dQ=TdS,\label{eq:dQ-1}
\end{equation}
where 
\begin{equation}
S=-k_{B}\sum_{n}P_{n}\ln P_{n}\label{eq:S}
\end{equation}
indicates the entropy of the particle, $k_{B}$ is Boltzmann's constant,
and
\begin{equation}
P_{n}=\exp\left(-\beta E_{n}\right)/\text{Tr}\left[\exp(-H/\left(k_{B}T\right))\right]\label{eq:pn}
\end{equation}
describes the occupation probability of a Gibbs state at energy $E_{n}$.
In the next section, the theory of fractional quantum mechanics and
the concepts of heat and work in quantum thermodynamic processes will
be applied to build quantum engines.

\section{\label{sec:QUANTUM-STIRLING-ENGINE}QUANTUM STIRLING ENGINE based
on fractional quantum mechanics}

Generally, the Stirling heat engine consists of two isothermal processes
and two isochoric processes\citep{chen1997effect,chen1998efficiency,wu1998performance,huang2014quantum}.
We focus on revealing the necessary conditions for the perfect regeneration
and the reversible operation based on fractional quantum mechanics.
For this reason, the fractional isothermal process, where the fractional
parameter and the well width are changed slowly, is proposed. This
process can be used to construct the fractional QSE, which consists
of two fractional isothermal processes $(A\rightarrow B$ and $C\rightarrow D)$
and two quantum isochoric processes $(B\rightarrow C$ and $D\rightarrow A)$,
as depicted in Fig. 1. The fractional parameter provides us with a
new way to regulate the thermodynamic cycle.

At stage I (A-B), the particle confined in the IPW interacts with
the hot bath at temperature $T_{h}$. The fractional parameter slowly
changes from $\alpha_{2}$ to $\alpha_{1}$ and the IPW varies from
$L_{A}$ to $L_{B}$. The process is infinitely slow, allowing the
particle to continually be in thermal equilibrium with the hot bath.
The probability of each eigenstate, which has the form of Eq. (\ref{eq:pn}),
changes from $P_{n}^{A}$ to $P_{n}^{B}$. With the help of Eq. (\ref{eq:dQ-1}),
the heat absorbed from the hot bath is written as
\begin{equation}
Q_{AB}=T_{h}[S(B)-S(A)],\label{eq:SQ-1}
\end{equation}
where $S(i)$ is the entropy of the particle at state $i$ calculated
by Eq. (\ref{eq:S}).

At stage II (B-C), the particle with the initial probability $P_{n}^{B}$
of each eigenstate is placed in contact with the the regenerator and
undergoes an isochoric process until reaching the temperature $T_{c}$.
The probability of each eigenstate changes from $P_{n}^{B}$ to $P_{n}^{C}$.
The eigenvalue $E_{n}$ of the fractional Hamiltonian operator $H$
is kept fixed as the well width and fractional parameter maintain
constant values, i.e., $L_{B}$ and $\alpha_{1}$, respectively. The
temperature of the particle decreases from $T_{h}$ to $T_{c}$. There
is heat exchange between the particle and the regenerator and no work
is performed in this isochoric process. According to Eq. (\ref{eq:dQ}),
the amount of the heat absorbed in this process is equal to the change
of the internal energy of the particle, i.e., 
\begin{equation}
Q_{BC}=U\left(C\right)-U\left(B\right)=\sum_{n}E_{n}\left(L_{B},\alpha_{1}\right)\left(P_{n}^{C}-P_{n}^{B}\right),
\end{equation}
where $U(i)$ is the internal energy of the particle at state $i$
calculated by Eq. (\ref{eq:U}). As $Q_{BC}<0$, heat is released
to the regenerator without any work being done.

At stage III (C-D), the particle is brought into contact with the
cold bath at temperature $T_{c}$. It is an isothermal process, which
is a reversed process of stage I. The state of the particle is always
in thermal equilibrium with the cold bath, while the fractional parameter
slowly changes from $\alpha_{1}$ to $\alpha_{2}$ and the IPW varies
from $L_{C}$ to $L_{D}$. Similar to Eq. (\ref{eq:SQ-1}), the heat
absorbed from the cold bath is 
\begin{equation}
Q_{CD}=T_{c}[S(D)-S(C)].\label{eq:SQ-1-1}
\end{equation}

At stage IV (D-A), the particle is removed from the cold bath and
goes through another isochoric process by connecting the the regenerator
until reaching the temperature $T_{h}$, where the well width and
fractional parameter are invariant. The cycle ends until the temperature
of the particle increasing to $T_{h}$. Heat absorbed from the regenerator
at this stage is computed by 
\begin{equation}
Q_{DA}=U\left(A\right)-U\left(D\right)=\sum_{n}E_{n}\left(L_{A},\alpha_{2}\right)\left(P_{n}^{A}-P_{n}^{D}\right).
\end{equation}
Note that $L_{D}=L_{A}$ is required for completing one cycle. 

\begin{figure}
\includegraphics{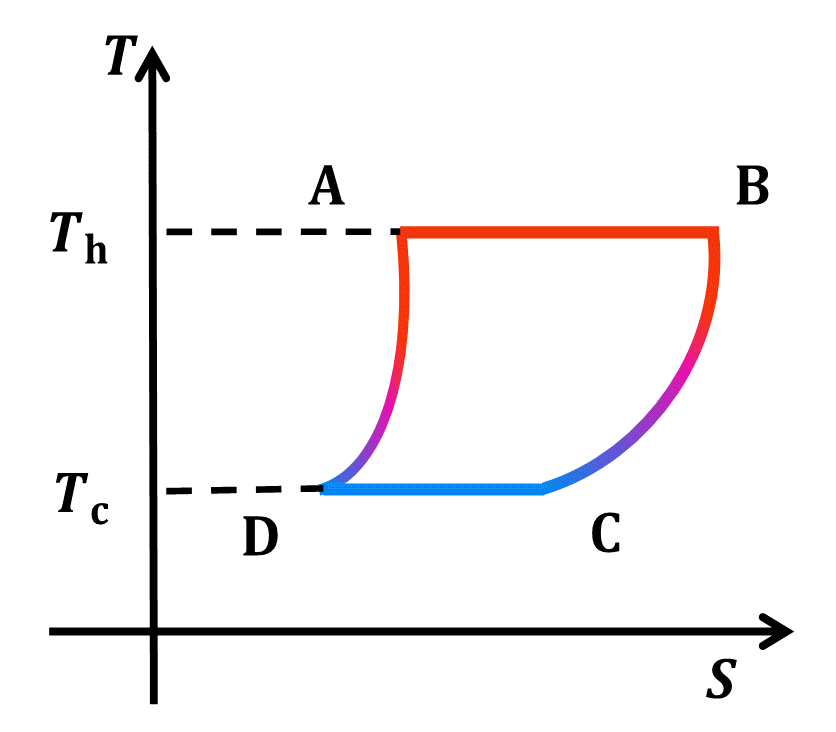}

\caption{Temperature-entropy (T-S) diagram for a quantum Stirling engine (QSE).}
\end{figure}
As the energy contained in the particle always returns to its initial
value. The net work done by the heat engine would then be

\begin{equation}
W=Q_{AB}+Q_{BC}+Q_{CD}+Q_{DA}.\label{eq:Ws}
\end{equation}

The Stirling heat engine is known as a closed-cycle regenerative heat
engine. The net heat exchange between the particle and the regenerator
during the two isochoric processes is 
\begin{equation}
Q_{R}=Q_{BC}+Q_{DA}.
\end{equation}
Three possible cases exists: (a) $Q_{R}=0$, (b) $Q_{R}<0$, and (c)
$Q_{R}>0$. The case $Q_{R}=0$ means that the regenerator is a perfect
regenerative heat exchanger. The mechanism of the perfect regeneration
makes the efficiency of the engine attain the Carnot value. When $Q_{R}\text{<}0$,
the heat $\left|Q_{BC}\right|$ flowing from the particle to the regenerator
in one regenerative process is larger than its counterpart $Q_{DA}$
flowing from the regenerator to the working substance in the other
regenerative process. The redundant heat in the regenerator per cycle
must be timely released to the cold bath. When $Q_{R}\text{>}0$,
the amount of $\left|Q_{BC}\right|$ is smaller than $Q_{DA}$. The
inadequate heat in the regenerator must be compensated from the hot
bath, otherwise the regenerator may not be operated normally. Due
to the non-perfect regenerative heat, the net heat absorbed from the
hot bath per cycle may be different from $Q_{h}$ and is given by
\begin{equation}
Q_{h}=Q_{AB}+H\left(Q_{R}\right)Q_{R},\label{eq:QH}
\end{equation}
where $H(x)$ is the Heaviside step function. The efficiency is an
important parameter for evaluating the performance, which is often
considered in the optimal design and theoretical analysis of heat
engines. 

By using Eqs. (\ref{eq:Ws}) and (\ref{eq:QH}), the expression of
the efficiency of the QSE should be

\begin{equation}
\eta=\frac{W}{Q_{h}}=\frac{W}{Q_{AB}+H\left(Q_{R}\right)Q_{R}}.\label{eq:effS}
\end{equation}

\section{\label{sec:result-and-discussion}resultS and discussion}

By using the model presented above, the performance of the QSE through
different ways of regulation will analyzed. Firstly, the QSE can be
regulated by adjusting the widths of the IPW for a given fractional
parameter value. Secondly, the fractional parameter can be adjusted
to identify the condition for perfect regeneration in the QSE when
the width of the IPW is fixed. Finally, the performance of the QSE
can be improved by simultaneously adjusting both the widths of the
IPW and the fractional parameters.

\subsection{THE EFFECTS OF WELL WIDTHS}

\begin{figure}
\includegraphics{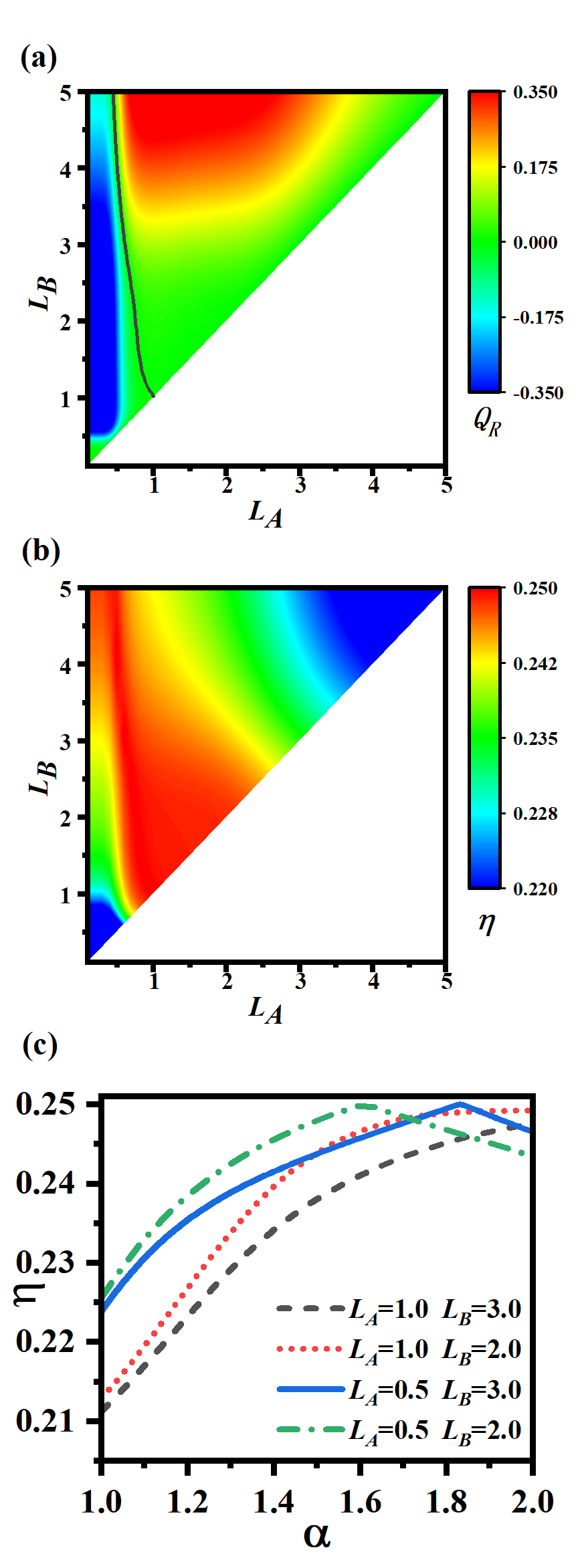}\caption{The contour plots of (a) the net heat exchange $Q_{R}$ between the
particle and the regenerator and (b) the efficiency $\eta$ varying
the widths $L_{A}$ and $L_{B}$ of the IPW, where $\alpha_{1}=\alpha_{2}=2$.
The black line represents the cycle with the perfect regeneration,
i.e., $Q_{R}=0$. (c) The efficiency $\eta$ of the Stirling cycle
as a function of the fractional parameter $\alpha$ for $L_{B}=L_{C}=2$
(dotted line and dash-dotted line) and $3$ (solid line and dashed
line), where $\alpha=\alpha_{1}=\alpha_{2}$, and $L_{A}=L_{D}=0.5$
and $1$, respectively. The parameters $T_{h}=4$, $T_{c}=3$, and
$m=1$. Note that Planck's constant $\hbar$ and Boltzmann's constant
$k_{B}$ are set to be unity throughout the paper, i.e., $\hbar=k_{B}=1$.}
\end{figure}

Fig. 2(a) shows the contour plot of the net heat exchange $Q_{R}$
between the particle and the regenerator of the QSE varying with the
widths $L_{A}$ and $L_{B}$ of the IPW, where the parameters $\alpha_{1}$
and $\alpha_{2}$ are set to be equal to 2. The optimizations of $L_{A}$
and $L_{B}$ yield the perfect regeneration with $Q_{R}=0$ {[}black
line in Fig. 2(a){]}. The contour plot of the efficiency $\eta$ of
the QSE as a function of $L_{A}$ and $L_{B}$ is presented in Fig.
2(b), and it can be observed that the region of Carnot efficiency
$\eta_{C}=1-T_{c}/T_{h}$ corresponds to that of perfect regeneration.

Fig. 2(c) shows the performance of a fractional QSE. In this case,
the fractional parameters $\alpha=\alpha_{1}=\alpha_{2}$ and the
well widths are some given values. The efficiency $\eta$ of the engine
is plotted as a function of the fractional parameter $\alpha$ for
different values $L_{B}=L_{C}=2$ (dotted and dash-dotted lines) and
$L_{B}=L_{C}=3$ (solid and dashed lines) of the well width, where
$L_{A}=L_{D}=0.5$ and $1$, respectively. The plot indicates that
when $L_{A}=L_{D}$ is about larger than $1$, the efficiency $\eta$
increases monotonically with $\alpha$ and reaches a maximum value
when $\alpha=2$ , which is the efficiency of the standard quantum
mechanical QSE. However, when $L_{A}=L_{D}$ is small, the efficiency
is not a monotonic function of $\alpha$. The optimal value of $\alpha$
can make the efficiency attain the Carnot efficiency. These results
mean that the performance of a QSE can be improved by regulating the
well widths and/or the fractional parameters.

\subsection{THE EFFECTS OF FRACTIONAL PARAMETERS}

\begin{figure}
\includegraphics{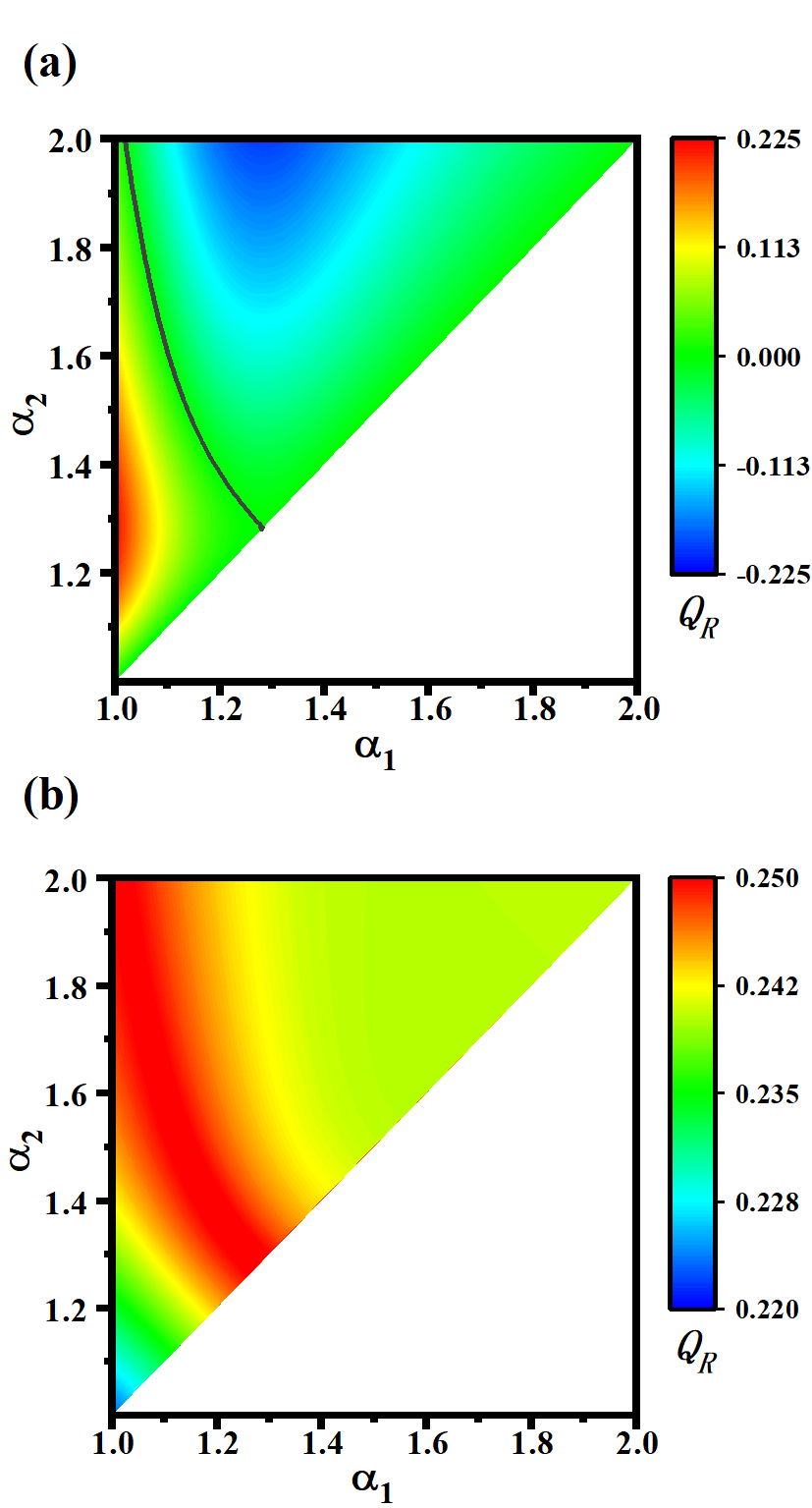}

\caption{The contour plots of (a) the net heat exchange $Q_{R}$ between the
particle and the regenerator and (b) the efficiency $\eta$ varying
the fractional parameters $\alpha_{1}$ and $\alpha_{2}$, where $T_{h}=4$,
$T_{c}=3$, $m=1$, and $L_{A}=L_{B}=L_{C}=L_{D}=1$. The black line
represents the cycle with the perfect regeneration, i.e., $Q_{R}=0$.}
\end{figure}

In this section, we examine the impact of regulating fractional parameters
on the performance of the QSE. The width of the IPW is kept constant
throughout the cycle, and the fractional parameter is slowly adjusted
from $\alpha_{2}$ ($\alpha_{1})$ to $\alpha_{1}$ ($\alpha_{2})$
during the fractional isothermal process from A to B (C to D), which
creates a QSE regulated solely by fractional parameters. To ensure
that the cycle proceeds forward, we set $\alpha_{1}<\alpha_{2}$.

By setting $L_{A}=L_{B}=L_{C}=L_{D}=1$ and combining Eqs.  (\ref{eq:SQ-1})-(\ref{eq:effS}),
the contour plot of the net heat exchange $Q_{R}$ between the particle
and the regenerator varying with $\alpha_{1}$ and $\alpha_{2}$ is
obtained, as shown in Fig. 3(a). The plot indicates that $Q_{R}$
is not a monotonic function of $\alpha_{1}$ and $\alpha_{2}$, and
the perfect regeneration is able to be achieved by optimizing these
parameters {[}black line in Fig. 3(a){]}. The contour plot of the
efficiency $\eta$ varying with $\alpha_{1}$ and $\alpha_{2}$ is
presented as well {[}see Fig. 3(b){]}. The plot shows that $\eta$
can reach the Carnot efficiency by optimizing $\alpha_{1}$ and $\alpha_{2}$.
This is because of the fact that suitable fractional parameters $\alpha_{1}$
and $\alpha_{2}$ lead to perfect regeneration $Q_{R}=0$.

\subsection{THE EFFECTS OF WELL WIDTHS AND FRACTIONAL PARAMETERS}

Fig.2 demonstrates that the QSE, which is controlled by the well widths,
does not achieve the optimal performance in most regions but can be
improved by introducing variational fractional parameters. To further
investigate this problem, we modify the isothermal process by adjusting
both the widths of the IPW and the fractional parameters simultaneously.
As an illustration, we consider the QSE with $L_{A}=1$ and $L_{B}=1.5$,
and shows how the engine\textquoteright s efficiency is enhanced by
the fractional parameters.

\begin{figure}
\includegraphics{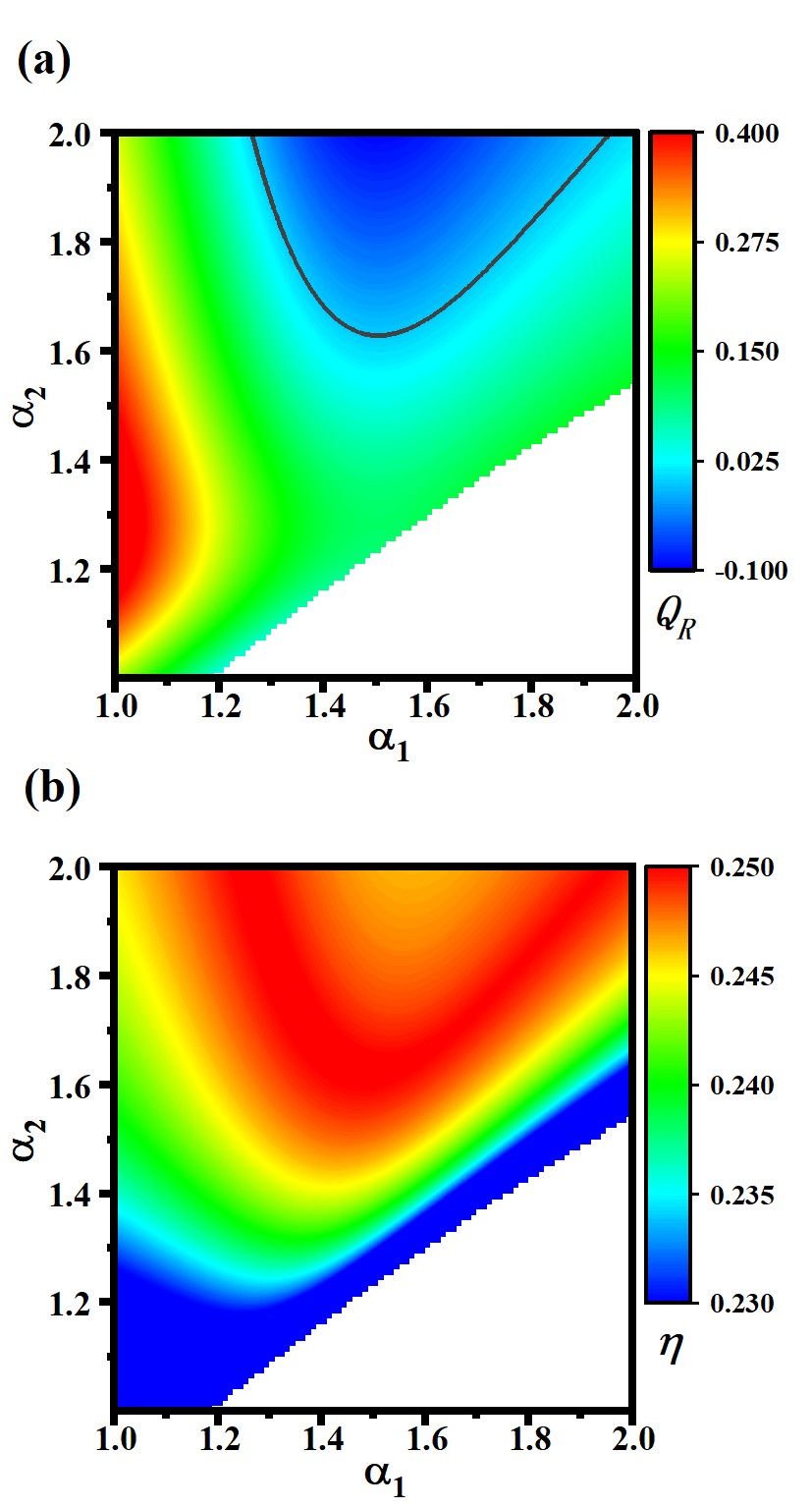}\caption{The contour plots of (a) the net heat exchange $Q_{R}$ between the
particle and the regenerator and (b) the efficiency $\eta$ varying
with the fractional parameters $\alpha_{1}$ and $\alpha_{2}$, where
$T_{h}=4$, $T_{c}=3$, $m=1$, $L_{A}=1$, and $L_{B}=1.5$. The
black line represents the cycle with the perfect regeneration, i.e.,
$Q_{R}=0$.}
\end{figure}

By combining Eqs.  (\ref{eq:SQ-1})-(\ref{eq:effS}), the contour
plot of the net heat exchange between the particle and the regenerator
$Q_{R}$ of the QSE varying with $\alpha_{1}$ and $\alpha_{2}$ is
provided {[}see Fig. 4(a){]}. It can be observed from the figure that
$Q_{R}$ is not a monotonic function of $\alpha_{1}$ and $\alpha_{2}$.
By optimizing $\alpha_{1}$ and $\alpha_{2}$, the cycle can achieve
perfect regeneration with $Q_{R}=0$. At the same time, the contour
plot of the efficiency $\eta$ varying with $\alpha_{1}$ and $\alpha_{2}$
is shown in Fig. 4(b). It can be observed from the figure that $\eta$
is also not a monotonic function of $\alpha_{1}$ and $\alpha_{2}$.
By optimizing $\alpha_{1}$ and $\alpha_{2}$, $\eta$ can reach the
Carnot efficiency. This indicates that the QSE solely regulated by
the widths of IPW may lead to a non-ideal regenerative cycle, but
the absolute value of the regenerative loss can be reduced and the
performance of the QSE can be improved by adjusting the fractional
parameters.

\begin{table}
\begin{tabular}{|c|c|c|c|c|}
\hline 
\multirow{2}{*}{$L_{A}$} & \multirow{2}{*}{$L_{B}$} & \multirow{2}{*}{$Q_{r}(\alpha_{1}=\alpha_{2}=2)$} & \multicolumn{2}{c|}{$Q_{r}=0$}\tabularnewline
\cline{4-5} \cline{5-5} 
 &  &  & $\alpha_{1}$ & $\alpha_{2}$\tabularnewline
\hline 
\multirow{2}{*}{0.6} & 0.9 & -0.1291 & 1.245 & 1.282\tabularnewline
\cline{2-5} \cline{3-5} \cline{4-5} \cline{5-5} 
 & 1.0 & -0.1315 & 1.279 & 1.326\tabularnewline
\hline 
\multirow{2}{*}{0.8} & 1.1 & -0.01223 & 1.311 & 1.409\tabularnewline
\cline{2-5} \cline{3-5} \cline{4-5} \cline{5-5} 
 & 1.2 & -0.01009 & 1.382 & 1.459\tabularnewline
\hline 
\multirow{2}{*}{1.0} & 1.3 & 0.005565 & 1.439 & 1.520\tabularnewline
\cline{2-5} \cline{3-5} \cline{4-5} \cline{5-5} 
 & 1.4 & 0.008296 & 1.502 & 1.579\tabularnewline
\hline 
\multirow{2}{*}{1.2} & 1.5 & 0.008021 & 1.517 & 1.621\tabularnewline
\cline{2-5} \cline{3-5} \cline{4-5} \cline{5-5} 
 & 1.6 & 0.01057 & 1.565 & 1.678\tabularnewline
\hline 
\multirow{2}{*}{1.4} & 1.7 & 0.007634 & 1.607 & 1.719\tabularnewline
\cline{2-5} \cline{3-5} \cline{4-5} \cline{5-5} 
 & 1.8 & 0.009979 & 1.660 & 1.778\tabularnewline
\hline 
\end{tabular}\caption{The values of fractional parameters $\alpha_{1}$ and $\alpha_{2}$
for the perfect regeneration at given values of the widths $L_{A}$
and $L_{B}$.}
\end{table}

\begin{figure}
\includegraphics{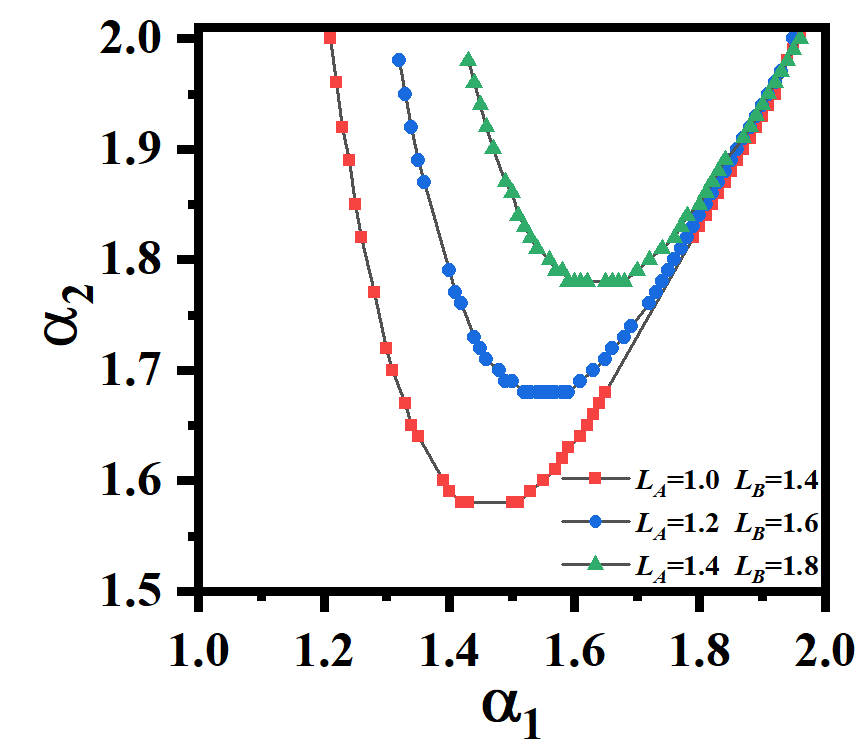}\caption{The fractional parameters $\alpha_{1}$ and $\alpha_{2}$ for the
perfect regeneration at $L_{A}=1.0,L_{B}=1.4$ (square points), $L_{A}=1.2,L_{B}=1.6$
(circular points), and $L_{A}=1.4,L_{B}=1.8$ (triangular points).}
\end{figure}

Furthermore, we demonstrate that by adjusting the fractional parameters,
the QSE with different well widths can achieve perfect regeneration
{[}see Table 1{]}. For given values of $L_{A}$ and $L_{B}$, the
third column of the table $1$ shows the regenerative loss $Q_{R}$
of the standard QSE ($\alpha_{1}=\alpha_{2}=2$), while the last two
columns show the optimal values of $\alpha_{1}$ and $\alpha_{2}$
for the cycle with perfect regeneration. In Fig. 5, we further present
the the fractional parameter $\alpha_{1}$ as a function of $\alpha_{2}$
under the condition of perfect regeneration for $L_{A}=1.0,L_{B}=1.4$
(square points), $L_{A}=1.2,L_{B}=1.6$ (circular points), and $L_{A}=1.4,L_{B}=1.8$
(triangular points). Fig.5 shows clearly that for different well widths,
the performance of the QSE can be improved through the regulation
of fractional parameters, and consequently, the Carnot efficiency
can be obtained.

\section{\label{sec:conclusion}conclusions}

By incorporating the fractional parameter into quantum thermodynamic
cycles, we have proposed a new way to regulate thermodynamic cycles
based on the fractional quantum mechanics. It is observed that the
energy level structure of the system can be changed by adjusting the
fractional parameters so that the perfect regeneration and the Carnot
efficiency are obtained. This proposal introduces a new approach for
designing thermodynamic cycles, when the motion of the particle transits
from Brownian motion to Lévy flight. Usually, Brownian motion is driven
by white Gaussian noise, whereas the Lévy process can be viewed as
a process driven by Lévy noise. Therefore, the introduction of fractional
quantum mechanics may provide us with a new route to study thermodynamic
processes that are affected by noise or some other heat engines with
specific properties. This may also allow us to investigate information
theory based on the fractional Schrödinger equation.
\begin{acknowledgments}
The authors thank Prof. Haijun Wang, Jia Du for helpful discussions
and comments. This work has been supported by the National Natural
Science Foundation (Grants No. 12075197) and the Fundamental Research
Fund for the Central Universities (No. 20720210024). 
\end{acknowledgments}

\appendix

\section*{APPENDIX A: \label{Appendix A}THE DERIVATION OF THE FRACTIONAL SCHRÖDINGER
EQUATION}

\setcounter{equation}{0} 
\global\long\def\theequation{A\arabic{equation}}%

During an infinitesimal interval $\varepsilon$, the state of the
fractional quantum-mechanical system evolves from $\psi(y,t)$ and
$\psi(x,t+\varepsilon)$, which is given by 

\begin{equation}
\psi(x,t+\varepsilon)=\int_{-\infty}^{\infty}dyK(x,t+\varepsilon\mid y,t)\psi(y,t).\label{eq:newphai}
\end{equation}

By using Eq. (\ref{eq:kernel}), the continuum limit $\sum_{j=1}^{N}V\left(x_{j}\right)\simeq\int_{t_{a}}^{t_{b}}d\tau V(x(\tau))$,
Feynman's approximation $\ensuremath{\int_{t}^{t+\varepsilon}d\tau V(x(\tau))\simeq\varepsilon V\left(\frac{x+y}{2}\right)}$,
and the kernel

\begin{equation}
\ensuremath{\begin{aligned} & K(x,t+\varepsilon\mid y,t)\approx\exp\left[-\frac{i}{\hbar}\varepsilon V\left(\frac{x+y}{2}\right)\right]\\
 & \times\lim_{N\rightarrow\infty}\int_{-\infty}^{\infty}dx_{1}\ldots dx_{N-1}\frac{1}{(2\pi\hbar)^{N}}\int_{-\infty}^{\infty}dp_{1}\ldots dp_{N}\\
 & \times\exp\left[\frac{i}{\hbar}\sum_{j=1}^{N}p_{j}\left(x_{j}-x_{j-1}\right)-\frac{i}{\hbar}D_{\alpha}\varepsilon\sum_{j=1}^{N}\left|p_{j}\right|^{\alpha}\right].
\end{aligned}
}\label{eq:KL}
\end{equation}
Note that 

\begin{equation}
\ensuremath{\sum_{j=1}^{N}p_{j}\left(x_{j}-x_{j-1}\right)=\sum_{j=1}^{N}x_{j}\left(p_{j}-p_{j+1}\right)+p_{N}x_{N}-p_{1}x_{0}}\label{eq:SUMp}
\end{equation}
and the $\delta$ function

\begin{equation}
\ensuremath{\delta\left(p_{j}-p_{j+1}\right)=\int dx_{j}\frac{1}{2\pi\hbar}\exp\left[\frac{i}{\hbar}x_{j}\left(p_{j}-p_{j+1}\right)\right]},\label{eq:deltafunction}
\end{equation}
Eq.  (\ref{eq:KL}) is simplified as

\begin{equation}
\begin{aligned}K_{L}(x,t+\varepsilon\mid y,t) & =\frac{1}{2\pi\hbar}\int_{-\infty}^{\infty}dp\exp\left[\frac{ip(x-y)}{\hbar}\right.\\
 & \left.-\frac{iD_{\alpha}|p|^{\alpha}\varepsilon}{\hbar}-\frac{i}{\hbar}\varepsilon V\left(\frac{x+y}{2},t\right)\right].
\end{aligned}
\label{eq:KLAP}
\end{equation}
Substituting Eq. (\ref{eq:KLAP}) into Eq. (\ref{eq:newphai}) arrives
at

\begin{equation}
\begin{aligned}\psi(x,t+\varepsilon)= & \int_{-\infty}^{\infty}dy\frac{1}{2\pi\hbar}\int_{-\infty}^{\infty}dp\\
 & \times\exp\left[\frac{ip(x-y)}{\hbar}\right]\times\exp\left[-\frac{i}{\hbar}D_{\alpha}|p|^{\alpha}\varepsilon\right]\\
 & \times\exp\left[-\frac{i}{\hbar}V\left(\frac{x+y}{2},t\right)\varepsilon\right]\psi(y,t).
\end{aligned}
\end{equation}

Expanding the left- and the right-hand sides in power series, taking
the first-order approximation and using the definition of Riesz operator
in Eq.  (\ref{eq:RieszD}), we have

\begin{equation}
\begin{aligned} & \psi(x,t)+\varepsilon\frac{\partial\psi(x,t)}{\partial t}\\
 & =\psi(x,t)+i\frac{D_{\alpha}\varepsilon}{\hbar}(\hbar\nabla)^{\alpha}\psi(x,t)-\frac{i}{\hbar}\varepsilon V(x,t)\psi(x,t),
\end{aligned}
\end{equation}
which can be further simplified to obtain Eq. (\ref{eq:FSeq}).

\bibliographystyle{apsrev4-1}
\bibliography{FQSE}

\end{document}